\documentclass[conference]{IEEEtran}
\IEEEoverridecommandlockouts
\usepackage{cite}
\usepackage{amsmath,amssymb,amsfonts}
\usepackage{algorithmic}
\usepackage{graphicx}
\usepackage{textcomp}
\usepackage{xcolor}
\usepackage{balance}
\def\BibTeX{{\rm B\kern-.05em{\sc i\kern-.025em b}\kern-.08em
    T\kern-.1667em\lower.7ex\hbox{E}\kern-.125emX}}

\usepackage[top=0.75in, bottom=1.05in, left=0.75in, right=0.75in]{geometry}

\usepackage{tikz}
\newcommand\copyrighttext{%
  \footnotesize \textcopyright 2026 IEEE. Personal use of this material is permitted. Permission from IEEE must be obtained for all other uses, in any current or future media, including reprinting/republishing this material for advertising or promotional purposes, creating new collective works, for resale or redistribution to servers or lists, or reuse of any copyrighted component of this work in other works.}
\newcommand\copyrightnotice{%
\begin{tikzpicture}[remember picture,overlay]
\node[anchor=south,yshift=10pt] at (current page.south) 
  {\fbox{\parbox{\dimexpr\textwidth-\fboxsep-\fboxrule\relax}{\copyrighttext}}};
\end{tikzpicture}%
}

\begin{document}
\bstctlcite{BSTcontrol}

\title{Toward LEO Satellite Network Systems for Instantaneous Detection of Environmental Changes\\
\thanks{We acknowledge the support of Research Manitoba, the Government of Canada, and the Natural Sciences and Engineering Research Council of Canada (NSERC), [funding reference number RGPIN-2022-03364]}
}

\author{
\IEEEauthorblockN{Zian Wang}
\IEEEauthorblockA{\textit{Center for Computational} \\
    \textit{Mathematics}\\
    {University of Waterloo}\\
Waterloo, Canada \\
zian.wang@uwaterloo.ca}
\and
\IEEEauthorblockN{Peng Hu}
\IEEEauthorblockA{\textit{Department of Electrical and} \\
    \textit{Computer Engineering}\\
{University of Manitoba}\\
Winnipeg, Canada \\
peng.hu@umanitoba.ca}
\and
\IEEEauthorblockN{Grant Gunn}
\IEEEauthorblockA{\textit{Department of Geography and} \\
    \textit{Environmental Management}\\
{University of Waterloo}\\
Waterloo, Canada \\
g2gunn@uwaterloo.ca}
}

\maketitle
\copyrightnotice
\begin{abstract}
The rapid deployment of Low Earth Orbit (LEO) satellite constellations has enabled the emergence of in-orbit edge computing and data centers, where satellites with onboard processing and high-speed inter-satellite links can collaboratively process data in space. This paper investigates whether such architectures, integrated with a deep learning–based computer vision pipeline, can achieve sub-minute information freshness suitable for real-time wildfire detection. To evaluate this hypothesis, we develop a simulation framework that models orbital dynamics, distributed processing, and network routing, using Age of Information (AoI) as the primary performance metric. A total of 720 simulation trials are conducted across 12 real-world constellation configurations, including Starlink, Kuiper, Telesat, and OneWeb. The results demonstrate that constellation design has a significant impact on AoI performance, with average AoI values ranging from 66.5 s to over 6300 s. The best-performing configurations achieve an average AoI below 70 s and a peak AoI under 100 s, indicating that orbital edge computing systems can provide the level of timeliness required for near-instantaneous environmental monitoring.
\end{abstract}


\section{Introduction}
The escalation of catastrophic wildfires globally has emerged as one of the most pressing environmental challenges of the 21st century. Forest fires impact twice the extent of tree cover compared to the early 2000s, with 2023 and 2024 marking some of the most destructive years recorded \cite{WRI_2025}. According to United Nations Environment Programme (UNEP), the frequency of extreme wildfire events is projected to increase by 30\% by 2050 and 50\% by the 2100 \cite{environment_spreading_2022}.

Historically, wildfire surveillance was fundamentally localized, relying on ground-based observation and human-centric monitoring. Traditional methods, such as direct visual assessment from fixed watchtowers and periodic airborne reconnaissance, provided high-fidelity in-situ data for specific, high-risk areas. However, traditional monitoring frameworks are inherently limited by their geographic scope\cite{chen_remote_2024}. While geostationary (GEO) platforms like the GOES-R series currently provide the frequent spaceborne updates for wildfire monitoring, their operational utility is fundamentally constrained by a critical resolution-frequency trade-off. As highlighted by \cite{zhao_goes-r_2022}, the GOES-R Advanced Baseline Imager (ABI) acquires images every 10 to 15 min, (while superior to the multi-day revisit times of traditional sun-synchronous satellites) but still imposes a significant lower bound on the Age of Information (AoI). In the context of rapidly spreading wildfires, a 15-min sampling interval represents a significant temporal bottleneck that can allow an ignition to transition into an uncontrollable event before the first alert is generated. Furthermore, satellites in a high orbit (e.g., GEO) have limited thermal spatial resolution, approximately 2 km per pixel, so early-stage ignitions may appear as sub-pixel hot spots that are difficult to detect reliably.

To address the limitations of traditional remote sensing approach, recent research has shifted toward Low Earth Orbit (LEO) architectures that may combine the high spatial resolution with the high temporal density of modern mega-constellations, although several challenges remain. Current LEO missions, such as Landsat 8, provide the 30-meter multispectral imagery required for precise fire segmentation but are limited by 16-day revisit cycles \cite{wulder_current_2019}. By contrast, emerging remote sensing constellations such as PlanetScope with hundreds of Dove satellites \cite{frassy_optical_2024}, or a LEO mega-constellation represented by Starlink can leverage large networks of interconnected nodes to potentially achieve much higher revisit rates. With high-speed inter-satellite links (ISLs) and significant onboard processing power, a LEO satellite constellation can function as a distributed orbiting computing cluster. This ``Orbital Edge Computing'' paradigm enables the local generation of fire masks, shifting the bottleneck from raw data transmission to distributed computational efficiency \cite{denby_orbital_2020}. This study is based on the hypothesis that, if each node in a mega-constellation were equipped with remote sensing capabilities comparable to Landsat 8 and dedicated multi-threaded computational resources, then shifting the bottleneck from data transmission to parallelized local computation could reduce the AoI to a level suitable for real-time wildfire suppression, thereby transforming the constellation into a reliable, high-fidelity pipeline. In this paper, we extend this paradigm through three main contributions:

\begin{itemize}
    \item Real-world LEO constellation configurations are employed (e.g., Starlink, Kuiper, Telesat, OneWeb) rather than hypothetical Walker-delta topologies.
    \item A specialized sensing task for wildfire detection is investigated, using a pixel-wise U-Net model on multi-spectral Landsat‑8 imagery over a considerably smaller region of interest that demands high coverage probability.
    \item Realistic processing latencies, derived from experiments on a high-performance computing (HPC) cluster, are incorporated into the analysis, grounding the simulations in observed computational delays.
\end{itemize}

Our contributions enable a rigorous evaluation of how physical architecture and orbital dynamics directly influence information freshness in time-critical environmental monitoring. The remainder of the paper is structured as follows: Section II discusses the related work; Section III presents the system model; Section IV discusses the experimental results; and Section V presents the conclusive remarks. 

\section{Related Work}

The development of real-time environmental monitoring systems in LEO draws from several evolving research domains. We categorize the existing literature into theoretical foundations, scheduling optimization, and edge applications.

\subsection{Theoretical Foundations of Information Freshness}
Recent literature has established theoretical foundations for data freshness by investigating the joint impact of sensing, transmission, and computation. An \textit{et al.} \cite{an_age_2024} derived closed-form expressions for the peak AoI in mobile edge computing (MEC) systems, identifying a ``U-shape'' performance trend where increasing the sensing rate initially improves freshness but eventually leads to queuing saturation. Their work emphasizes that a partial computation strategy—distributing tasks between local and edge servers—offers superior stability and lower peak AoI compared to purely local or edge-based processing. Theoretical analysis by Tang \textit{et al.} \cite{tang_age_2021} provides closed-form expressions for average AoI in multi-user settings, revealing that while edge computing minimizes AoI for small user groups, local computing becomes more effective as network congestion increases.

\subsection{Optimization and Scheduling Frameworks}
Beyond AoI models, research has been conducted toward optimizing the complex task flows of satellite networks were conducted. A distributed deep reinforcement learning (DRL) approach \cite{li_delay-cost_2025} was proposed to minimize delay costs by predicting fluctuating satellite loads and optimizing multi-hop inter-satellite offloading. Addressing the integration of these tasks, Jaipuria \textit{et al.} \cite{jaipuria_distributed_2025} proposed an integrated collect-communicate-compute framework for LEO constellations. By utilizing an integer linear programming (ILP) based solution, this approach generates optimal schedules that maximize the number of regions collected and analyzed.

\subsection{Onboard Edge Intelligence and Applications}
The practical feasibility of these paradigms is increasingly demonstrated through targeted applications and hardware benchmarking. The specific application of distributed edge computing architectures to LEO constellations was introduced for vessel detection using YOLOv8 on simplified Walker-delta shells \cite{vessel_detection}. The feasibility of such pipelines has been further strengthened by testing deep learning (DL) models, such as VFNet and RADCNet, for maritime identification on representative edge-AI hardware, including Raspberry Pi and VPU-based architectures. An onboard edge AI approach for LEO satellites \cite{ZhangW25, ZhangW24} is proposed for detecting space objects and supporting space sustainability. These studies demonstrate that end-to-end processing of raw imagery can be achieved within stringent power budgets and latency constraints \cite{del_prete_enhancing_2025}. Furthermore, transfer learning approaches are being utilized to enhance the performance of lightweight models for environmental tasks; for instance, Gain \textit{et al.} \cite{gain_leo_2024} demonstrated that a low-parameterized MobileNetV2 model can be fine-tuned using transfer learning to achieve near 100\% accuracy in wildfire detection.

Based on the aforementioned works, we can see the theoretical foundations, scheduling optimization, and practical edge applications research domains are largely built on the assumption of hypothetical satellite networks with idealized computational power and transmission capability. The need exists to identify the capacity of current LEO network system models to reduce AoI for circumstances where immediate classification of imagery and transmission is needed. Our paper addresses this important gap in the current literature. 

\section{System Model}

\subsection{System Overview}
We consider a LEO satellite constellation \(\mathcal{S}\) comprising \(P\) orbital planes, each with \(R\) satellites, deployed at altitude \(h\) and inclination \(\iota\). The constellation follows a Walker pattern (\(\Delta\) or \(\star\)) with phasing parameter \(F\). A single ground station \(g\) with geodetic coordinates \((\phi_g,\lambda_g)\) serves as the sink for all processed data. A fixed geographic region of interest (RoI) \(\mathcal{R}\) (e.g., a wildfire-prone area) is monitored continuously.

The distributed edge computing workflow proceeds in four stages. First, a satellite that has the RoI entirely within its sensor swath is designated as the \emph{master node}. It initiates a sensing task by capturing a raw multispectral image. Second, the master partitions the image into smaller patches using the Geospatial Data Abstraction Library (GDAL), distributing these patches to itself and a set of neighboring worker satellites. Third, each worker executes a pretrained U-Net-based active fire detection model \cite{de_almeida_pereira_active_2021} on its assigned patches in parallel, generating binary fire masks. Finally, the resulting masks are packetized and routed through the time-varying ISL network toward the ground station using a shortest-path strategy. The system is evaluated using the AoI metric, which jointly captures the freshness of the received fire alerts.

\subsection{Age of Information}
AoI is used as the metric to quantify the timeliness of wildfire alerts. At any time \(t\), the AoI is defined as
\begin{equation}
AoI(t) = t - I(t),
\label{eq:aoi}
\end{equation}
where \(I(t)\) is the generation time of the most recent update received at the ground station by time \(t\). Between updates, \(AoI(t)\) increases linearly with slope 1. When a new update with generation time \(t_n\) is fully received at time \(t_n'\), the AoI resets to \(t_n' - t_n\) if this value is smaller than the current AoI; otherwise, the update is discarded as obsolete. This mechanism yields the classic sawtooth pattern of AoI.

We evaluate system performance over a finite simulation horizon of length \(T\). The time is discretized into steps of size \(\Delta t\), yielding timestamps \(t_0, t_1, \dots, t_N\). The average AoI is then computed as
\begin{equation}
\overline{AoI} = \frac{1}{N+1} \sum_{i=0}^{N} AoI(t_i).
\label{eq:aavg}
\end{equation}
In addition, the peak AoI is defined as \(AoI_{\text{peak}} = \max_{0 \le i \le N} AoI(t_i)\), which captures the worst-case staleness and is critical for applications with hard timeliness constraints.

\subsection{Orbital and Network Model}

\begin{figure}
    \centering
    \includegraphics[width=0.32\textwidth]{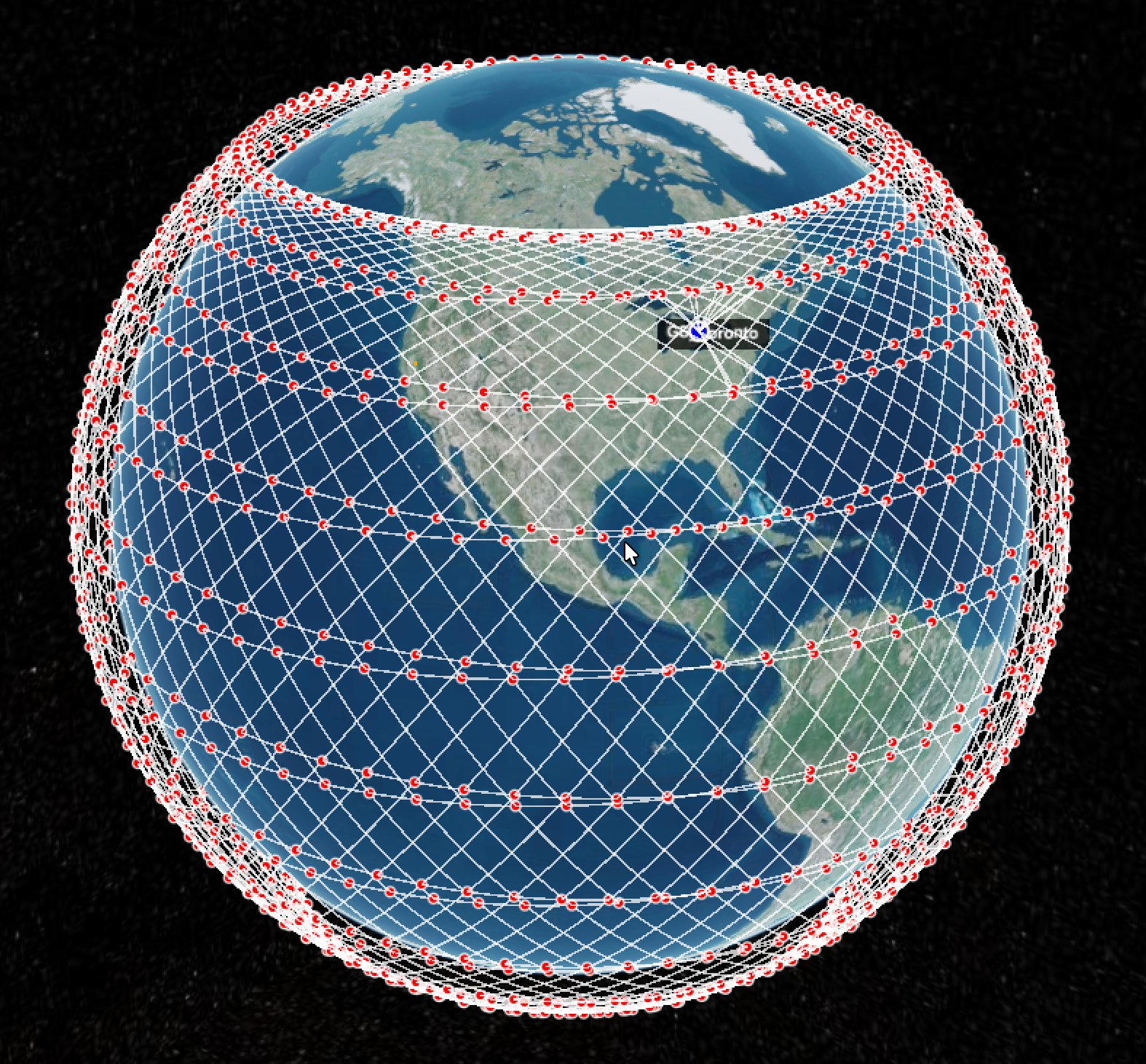}
    \caption{Topology visualization for one snapshot of Starlink s1}
    \label{fig:grid_plus}
\end{figure}

Each constellation is defined by the tuple \(\mathcal{C} = \langle P,R,\iota,h,\theta_{\min},\mathcal{W},F\rangle\), where \(P\) orbital planes each contain \(R\) satellites at altitude \(h\), inclination \(\iota\), and minimum satellite-to-ground elevation threshold \(\theta_{\min}\). The Walker pattern \(\mathcal{W}\) (Delta or Star) and phasing \(F\) determine the relative spacing. Satellites follow circular orbits; their positions are propagated using Keplerian dynamics and transformed to the Earth-Centered Earth-Fixed (ECEF) frame via standard orbital mechanics. The communication network at time \(t_i\) is a directed graph \(G(t_i)=(\mathcal{V},\mathcal{E}(t_i))\) with vertices \(\mathcal{V}=\mathcal{S}\cup\{g\}\) comprising all satellites \(\mathcal{S}\) and one ground station \(g\). Edges include intra-plane ISLs to the two neighbors in the same orbital plane, inter-plane ISLs to adjacent planes (with seam‑closing using phasing \(F\)), and satellite-to-ground links (SGLs) whenever the elevation angle from the ground station to a satellite exceeds \(\theta_{\min}\). This topology follows the Grid+ design \cite{mclaughlin_grid_2023}, Each satellite maintains up to four ISL neighbors (i.e., two intra‑plane, two inter‑plane), providing multiple redundant paths for routing, as depicted in Fig.~\ref{fig:grid_plus}.

A satellite \(s\) becomes an active sensing node (i.e., \(s\in\mathcal{A}(t_i)\)) only if the entire RoI \(\mathcal{R}\) lies within its sensor swath. Using an azimuthal equidistant projection centered at the sub‑satellite point, the swath is modeled as a circle of radius \(R_{\text{swath}}\); the satellite is active if the projected vertices of \(\mathcal{R}\) are all inside this circle. This ensures that any captured scene fully contains the target area.

\subsection{Distributed Processing and Data Plane}
When a satellite \(s\) enters the active sensing set \(\mathcal{A}(t_i)\), it first incurs a preprocessing delay \(\delta_{\text{pre}}\) to decompress and partition the captured raw image \(\mathcal{M}_{\text{raw}}\) into patches using GDAL. It then distributes these patches among itself and a set of worker nodes \(\mathcal{W}_s \subseteq \text{Neighbors}(s,G(t_i))\). Each worker executes a U‑Net inference pipeline on its assigned patches, producing binary fire masks \(\mathcal{M}_{\text{proc}}\). The effective computation delay is modeled as \(\Delta_{\text{comp}} = \delta_{\text{inf}} / (1 + |\mathcal{W}_s|)\), where \(\delta_{\text{inf}}\) is the single‑node inference time, reflecting parallel speedup.

The resulting masks are packetized into fixed-size packets of \(\omega\) bits, where \(\omega\) denotes the maximum transmission unit (MTU). To emulate real network operation while maintaining simulation efficiency, \(\omega\) is set to the largest IP datagram size. Each packet is independently injected into the network and routed toward the ground station \(g\) using a shortest-path strategy based on the current topology \(G(t_i)\). Links are modeled as \textit{SimPy} resources with capacity 1, and a packet must acquire the link resource before transmission, incurring queuing delay \(\delta_{\text{queue}}\). The hop latency for a packet traversing edge \((u,v)\) is
\[
L_{uv} = \delta_{\text{queue}}(u,v) + \frac{\omega}{B_{uv}} + \frac{d(u,v)}{c},
\]
where \(B_{uv}\) is the link bandwidth (e.g., 10 Gbps for ISLs, 100 Mbps for SGLs), \(d(u,v)\) the Euclidean distance, and \(c\) the speed of light. When the constellation state updates from \(G(t_i)\) to \(G(t_{i+1})\), any in‑flight packets are rerouted using the new topology. A task is considered complete only when all packets carrying the fire mask for a given scene are received and reassembled at the ground station.

\section{Experimental Results and Analysis}
To evaluate the proposed distributed sensing framework, we first characterize computational delays that govern the on-board processing pipeline. These delays were realistically measured on the HPC cluster Nibi using an Landsat‑8 scene (Product ID LC08\_L1TP\_046031\_20200908\_20200908\_01\_RT) \cite{landsat8_lc08_l1tp_046031_20200908}. The node was equipped with 8 CPU cores, 32GB RAM, and an NVIDIA H100 GPU configured as a 3g.40gb Multi‑Instance GPU (MIG) slice to emulate the partitioned computing resources typical of satellite edge nodes. The GDAL pre-processing step, which splits the \(7811\times7931\) pixel scene into \(256\times256\) patches, was executed 100 times, yielding an average delay \(\delta_{\text{pre}} = 49.62\) s (\(\sigma = 12.28\) s). The U‑Net inference pipeline was similarly run 100 times, producing an average inference time \(\delta_{\text{inf}} = 43.27\) s (\(\sigma = 9.00\) s). These measured latencies serve as the basis for the computational delays in the simulation model.

The experiment matrix is defined by the Cartesian product of 12 constellation shells, 4 ground stations, 5 swath radii, and 3 simulation time intervals. This section reports performance across these configurations, with a focus on how constellation design, swath geometry, and temporal resolution jointly determine AoI.

\subsection{AoI Performance Across Constellation Shells}
To evaluate the impact of orbital design on AoI, Table~\ref{tab:aoi_comparison} compares all 12 constellation shells at a swath radius of 500 km, averaged across 4 ground stations and 3 simulation time intervals.

\begin{table}[!ht]
\centering
\caption{Comparison of AoI Across Constellation Shells ($\text{Swath radius} = 500$ km). Results are averaged across 4 ground stations and 3 simulation timesteps (10s, 20s, 30s) over a 24-hour mission window.}
\label{tab:aoi_comparison}
\begin{tabular}{l c c}
\hline
\textbf{Constellation Shell} & \textbf{Average AoI (s)} & \textbf{Peak AoI (s)} \\ \hline
\textit{Starlink} & & \\
starlink\_s1 & 67.1 & 99.3 \\
starlink\_s2 & 268.9 & 1,419.3 \\
starlink\_s3 & 2,697.0 & 10,358.0 \\
starlink\_s4 & 66.8 & 99.3 \\
starlink\_s5 & 6,318.6 & 19,440.0 \\ \hline
\textit{Amazon Kuiper} & & \\
kuiper\_k1 & 66.5 & 76.0 \\
kuiper\_k2 & 3,914.6 & 12,971.0 \\ \hline
\textit{Telesat} & & \\
telesat\_t1 & 192.2 & 456.0 \\
telesat\_t2 & 66.5 & 76.0 \\ \hline
\textit{OneWeb} & & \\
oneweb\_o1 & 66.5 & 78.0 \\
oneweb\_o2 & 156.5 & 1,895.8 \\
oneweb\_o3 & 704.0 & 4,394.2 \\ \hline
\end{tabular}
\end{table}

\begin{figure}
    \centering
    \includegraphics[width=0.45\textwidth]{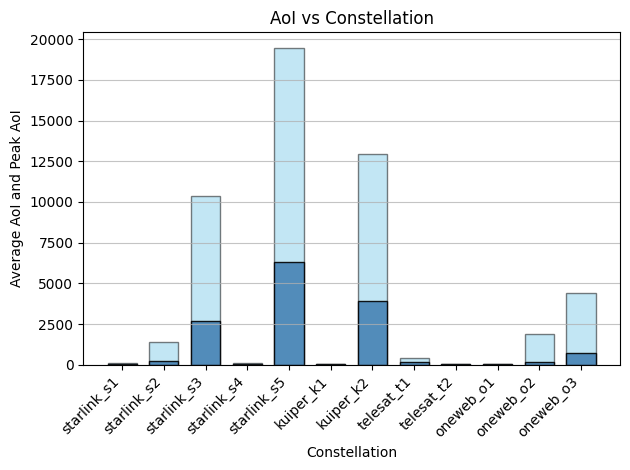}
    \caption{AoI performance comparison across constellation shells at a swath radius of 500 km. Average AoI is shown in dark blue.}
    \label{fig:aoi_comparison}
\end{figure}

Also as shown in Fig. \ref{fig:aoi_comparison}, a substantial variation on the AoI performance is observed across different constellation shells, with the average AoI ranging from 66.5 s to 6318.6 s. The best performing constellation shells (starlink\_s1, starlink\_s4, kuiper\_k1, telesat\_t2, oneweb\_o1) all have an average AoI of approximately 66.5 s, while the worst performing constellation shell (starlink\_s5) has an average AoI of 6318.6 s, which is almost 100 times greater than the best performing shells.

\begin{figure}[!ht]
    \centering
    \includegraphics[width=0.45\textwidth]{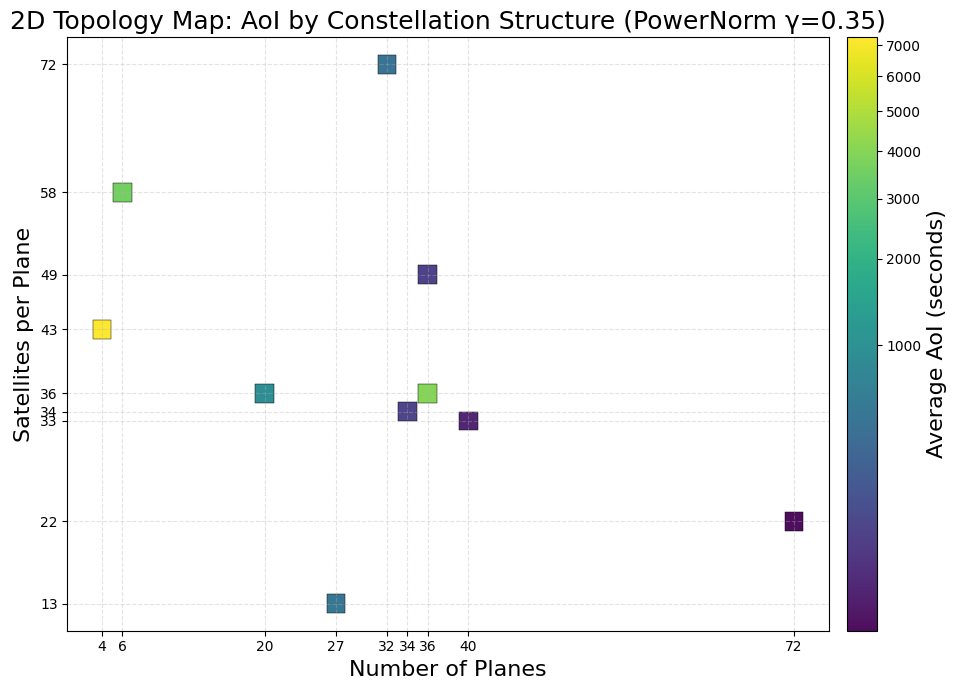}
    \caption{AoI performance topology as a function of constellation parameters $P$ and $R$. Each point represents a unique constellation shell configuration, with color indicating the average AoI (darker colors indicate higher AoI).}
    \label{fig:P_against_R}
\end{figure}

To further understand the reasons different constellations affect average AoI, we analyzed the relationship between the $P$, $R$ and the resulting average AoI. By isolating the 10 unique configurations representing current major real-world constellation, we obtain a 2D topology map of average AoI performance as a function of $P$ and $R$ as shown in Fig. \ref{fig:P_against_R}.

Under the current simulation setting, the topology pattern indicates that increasing the number of orbital planes ($P$) has a stronger effect in reducing Average AoI than simply increasing the number of satellites per plane ($R$). Configurations with larger $P$ tend to cluster at lower Average AoI levels even when $R$ is moderate.

\subsection{Average AOI performance against swath radius}

\begin{figure}[!ht]
    \centering
    \includegraphics[width=0.48\textwidth]{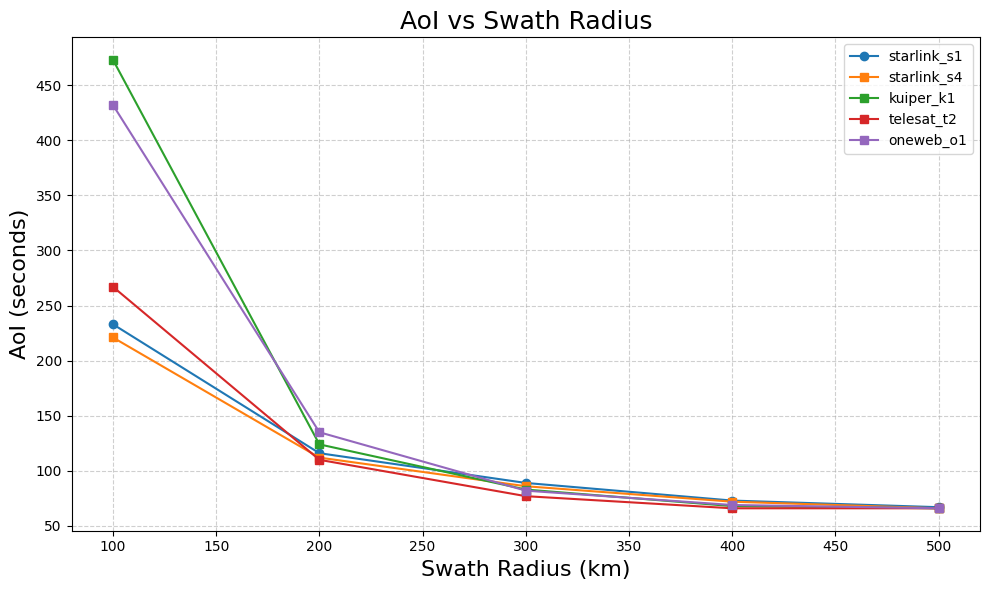}
    \caption{Average AOI against swath radius for selected shells}
    \label{fig:good_shell_swath}
\end{figure}

To evaluate the impact of swath radius on average AoI performance, we compare representative constellation shells across five swath radii (100, 200, 300, 400, and 500 km), averaged over four ground stations and three simulation intervals as shown in Table~\ref{tab:swath_aoi_comparison} and Fig. \ref{fig:good_shell_swath}. These shells are selected because, at large swath radii, their average AoI approaches the processing-time floor of the system.

\begin{table}[!ht]
\centering
\caption{Impact of Swath Radius on Average AOI (in seconds) across Representative Constellation Shells.}
\label{tab:swath_aoi_comparison}
\resizebox{0.45\textwidth}{!}{
    \begin{tabular}{l c c c c c}
    \hline
    \textbf{Constellation Shell} & \textbf{100 km} & \textbf{200 km} & \textbf{300 km} & \textbf{400 km} & \textbf{500 km} \\ \hline
    starlink\_s1 & 233.0 & 116.0 & 89.0 & 73.0 & 67.0 \\
    starlink\_s4 & 221.0 & 112.0 & 86.0 & 72.0 & 66.0 \\
    kuiper\_k1 & 473.0 & 124.0 & 83.0 & 68.0 & 66.0 \\
    telesat\_t2 & 267.0 & 110.0 & 77.0 & 66.0 & 66.0 \\
    oneweb\_o1 & 432.0 & 135.0 & 82.0 & 69.0 & 66.0 \\ \hline
    \end{tabular}
}

\end{table}

For these selected configurations, the average AoI converges toward an asymptotic floor of approximately 66--67 s. This threshold corresponds closely to the cumulative latency of the image pre-processing, inference phases, and time interval overhead. This confirms that for these specific orbital shells, the average AoI performance is no longer limited by the satellite revisit interval after 400 km or 500 km, but rather by the on-board computational and data-handling delay. Conversely, at a constrained swath radius of 100 km, the average AoI performance is significantly degraded even for those selected constellation shells, with an average ranging from 221 to 473 s across the selected shells. This disparity indicates that at a smaller swath radius, the information freshness is primarily dictated by orbital dynamics. Moreover, the relationship between swath radius and average AoI is characterized by a distinct non-linear decay. While the swath radius increases linearly, the corresponding reduction in average AoI exhibits diminishing marginal returns. The most substantial performance gains are realized between 100 km and 300 km. Beyond this point, the curves begin to flatten considerably.

\subsection{AoI performance against time intervals}
A critical factor affecting the timeliness of fire detection data is the frequency at which sensing tasks are initiated within the simulation. This frequency, ultimately adjusted by the simulation time interval ($\Delta t$), determines how often the constellation identifies new sensing opportunities and updates its network routing state. Analyzing this parameter is essential to understand whether the current system performance is limited by the physical network capacity or by the granularity of the task scheduling itself.

\begin{figure}[!ht]
    \centering
    \includegraphics[width=0.48\textwidth]{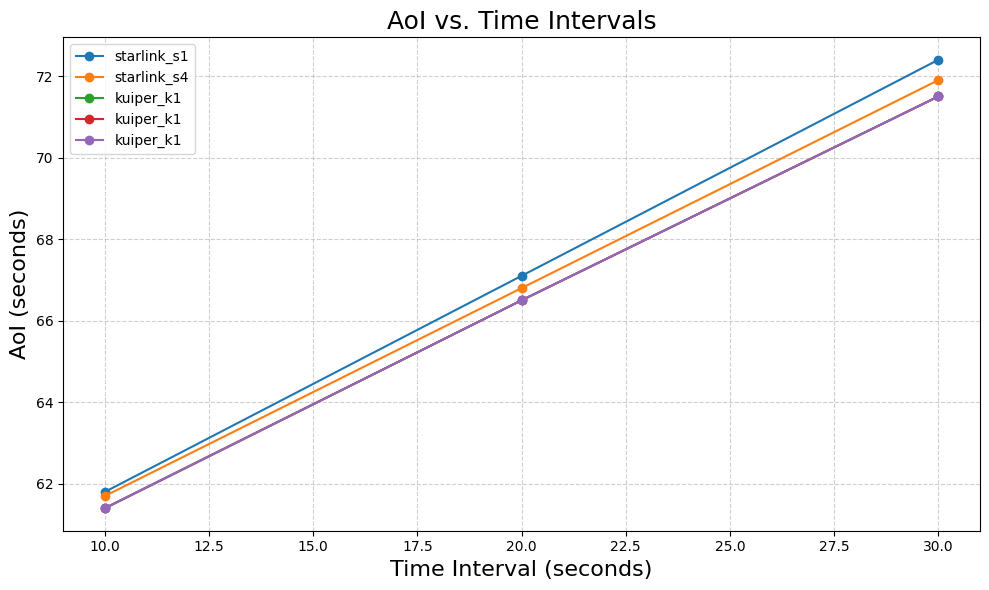}
    \caption{AoI performance against time intervals}
    \label{fig:time_interval_aoi}\vspace{-10pt}
\end{figure}

As shown in Table \ref{tab:time_interval_aoi} and Fig. \ref{fig:time_interval_aoi}, the simulation results across high-density shells demonstrate a nearly linear relationship between the reduction in task intervals and the improvement in average AoI. Across these tested intervals, average AoI decreases consistently as the task interval is reduced from 30 s to 10 s for all selected shells.

\begin{table}[htp]
\centering
\caption{Sensitivity of average AoI (in seconds) to Simulation Time Intervals ($R_{swath} = 500$ km).}
\label{tab:time_interval_aoi}
\begin{tabular}{l c c c}
\hline
\textbf{Constellation Shell} & \textbf{30s Interval} & \textbf{20s Interval} & \textbf{10s Interval} \\ \hline
starlink\_s1 & 74.0 & 67.3 & 60.1 \\
starlink\_s4 & 73.1 & 66.8 & 59.8 \\
kuiper\_k1   & 73.0 & 66.5 & 59.5 \\
telesat\_t2  & 72.9 & 66.5 & 59.5 \\
oneweb\_o1   & 72.8 & 66.5 & 59.4 \\ \hline
\end{tabular}
\end{table} \vspace{-10pt}

\subsection{Coverage Probability and AoI}
\begin{figure}[!ht]
    \centering
    \includegraphics[width=0.48\textwidth]{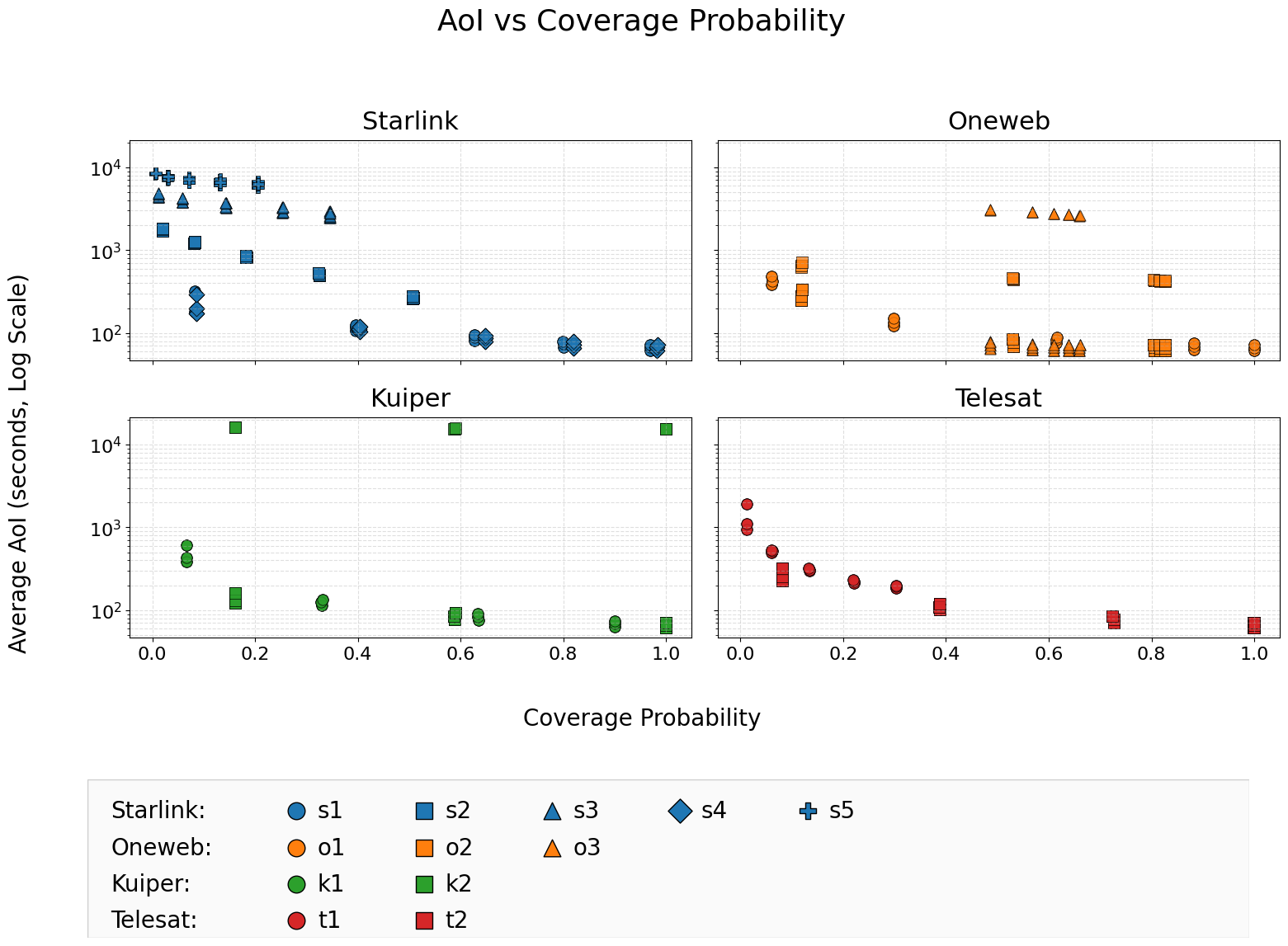}
    \caption{AoI performance vs. coverage probability}
    \label{fig:coverage_vs_aoi} \vspace{-10pt}
\end{figure}

While the internal network bandwidth is not a primary bot-tleneck, the system’s performance will be heavily influenced by the temporal resolution of the sensing tasks. This suggests that the timeliness of a fire detection update is inextricably linked to how frequently the constellation can geographically observe the target area. To quantify this relationship, the Coverage Probability is evaluated, which is a metric representing the percentage of total simulation time intervals in which at least one satellite in the constellation has the RoI within its sensor swath. From Fig. \ref{fig:coverage_vs_aoi}, with our 24-hour simulation window, the Coverage Probability is calculated as the ratio of ``active'' intervals (where the RoI is covered) to the total number of simulation steps. To analyze the impact of this temporal visibility, a cross-constellation comparison was performed, mapping this probability against the resulting average AoI for all 720 simulation trials.

Categorized by satellite family and individual orbital shells in revealing an inverse relationship. As the percentage of intervals with active coverage increases toward 100\%, the average AoI drops across all constellation families. Another key observation is the behavior of the data as it reaches 100\% coverage probability. For all data points where the target is visible in every simulation interval, there is a significantly smaller spread in the average AoI values compared to lower coverage levels. This reduction in variance marks a fundamental shift in system behavior as the elimination of revisit interval gaps removes the primary source of uncertainty. At this stage, the system transitions from a stochastic state to a more deterministic one, where the performance is no longer a game of chance based on satellite positions but is instead governed by the fixed, predictable latencies of the hardware, network, and time interval.

\section{Conclusion}
This paper presents a simulated orbital edge-computing architecture embedded within realistic LEO constellations, demonstrating its ability to maintain information freshness at levels suitable for time-critical wildfire monitoring. The proposed framework combines orbital dynamics, distributed U-Net inference, and time-varying ISL routing, with computation delays grounded in empirical measurements. The results highlight the dominant role of constellation design in governing AoI performance. Across 12 orbital shells, average AoI ranges from approximately 66.5 s to over 6300 s, with the most effective configurations achieving sub-70 s mean and sub-100 s peak AoI. Notably, increasing the number of orbital planes and enhancing coverage probability have a more pronounced impact on timeliness than simply adding satellites per plane. Once near-continuous coverage is attained, AoI becomes primarily constrained by onboard processing and scheduling granularity rather than communication bandwidth, while the consistent improvements observed with shorter task intervals suggest that ISL capacity remains underutilized within the evaluated regime. Overall, these findings support the feasibility of LEO-based distributed sensing for near-real-time environmental alerting and motivate future work on multi-ground-station networks, heterogeneous onboard computing resources, and adaptive task scheduling under realistic operational constraints.

\balance
\bibliographystyle{IEEEtran}
\bibliography{references.bib}   

\end{document}